\documentclass{article}

\usepackage{PRIMEarxiv}

\usepackage[utf8]{inputenc} 
\usepackage[T1]{fontenc}    
\usepackage{hyperref}       
\usepackage{url}            
\usepackage{booktabs}       
\usepackage{amsfonts}       
\usepackage{nicefrac}       
\usepackage{microtype}      
\usepackage{lipsum}
\usepackage{fancyhdr}       
\usepackage{graphicx}       
\graphicspath{{media/}}     

\usepackage{natbib}
\usepackage{subcaption}
\usepackage{xcolor}

\title{Deception Analysis with Artificial Intelligence: An Interdisciplinary Perspective
\thanks{Ștefan Sarkadi. Deception Analysis with Artificial Intelligence: An Interdisciplinary Perspective. (Work in Progress)} 
}

\author{
  Ștefan Sarkadi \\
  Dept. of Informatics \\
  King's College London \\
  London, UK\\
  \texttt{stefan.sarkadi@ekcl.ac.uk} \\
}

\begin{document}
\maketitle

\begin{abstract}
Humans and machines interact more frequently than ever and our societies are becoming increasingly hybrid. A consequence of this hybridisation is the degradation of societal trust due to the prevalence of AI-enabled deception. Yet, despite our understanding of the role of trust in AI in the recent years, we still do not have a computational theory to be able to fully understand and explain the role deception plays in this context. This is a problem because while our ability to explain deception in hybrid societies is delayed, the design of AI agents may keep advancing towards fully autonomous deceptive machines, which would pose new challenges to dealing with deception. In this paper we build a timely and meaningful interdisciplinary perspective on deceptive AI and reinforce a 20 year old socio-cognitive perspective on trust and deception, by proposing the development of \textit{DAMAS} - a holistic Multi-Agent Systems (MAS) framework for the socio-cognitive modelling and analysis of deception. In a nutshell this paper covers the topic of modelling and explaining deception using AI approaches from the perspectives of Computer Science, Philosophy, Psychology, Ethics, and Intelligence Analysis. 
\end{abstract}

\keywords{Deception Analysis \and Deceptive AI \and Intelligence Analysis \and Argumentation \and Multi-Agent Systems}

\section{Introduction}


History, Economics, Politics, Philosophy, Communication Sciences, Sociology, and the Cognitive Sciences have looked at deception from perspectives that are predominantly anthropocentric. Thus, the significant knowledge we have about deception revolves around its human nature. This acquired knowledge emphasises that deception plays an important role for humans and that deception is a multi-layered phenomenon which takes numerous forms during social interactions.

However, more recently, the anthropocentric grip on understanding deception has weakened. Research on deception (and its detection) is expanding beyond human agents, to deceptive technologies, due to the current hybridisation of our societies. Hybrid societies are \textit{`self-organizing, collective systems, which are composed of different components, for example, natural and artificial parts (bio-hybrid) or human beings interacting with and through technical systems (socio-technical)'} \citep{hamann2016hybrid}. Nowadays, AI technologies play a crucial role in hybrid societies, but research in AI and deception has not progressed enough to allow us to understand and predict how advancements in the design of AI agents will impact hybrid societies. 

A particular threat to the hybridisation of societies is the development of fully autonomous deceptive AI agents that will be able to form their own reasons and methods to perform deception, as well as out-think and outsmart humans and other AI agents \citep{sarkadi2020phdthesis}. By fully autonomous deceptive AI agent we mean neither the already existing human-scripted `mindless' chatterbots which follow a pre-programmed script to deceive \citep{mauldin1994chatterbots}, nor the `clueless' stochastic parrots \citep{bender2021dangers} which blurt out sentences without having any sense of their meaning in-context, but AI agents in the likes of the conceptual machines that trick the judges in the Imitation Game \citep{Turing1950}. In particular, we mean AI agents in the likes of the ones formally described by \citep{cohen1985speech} or in the likes of Hamblin machines \citep{staines2018linguistics}, which, in addition to being able to reason about what they communicate to others in various contexts and situations and how their communicative actions cause changes in the minds of others, also have deceptive motives.

To make informed decisions regarding the development and deployment of deceptive AI agents we need to understand how they interact with humans and with each other. To do this, we can computationally model hybrid societies and observe their cognitive processes and emergent behaviour. The focus of this paper is precisely on the modelling of artificial deceptive agents inside societies to study and explain potential outcomes.

\begin{figure}
     \centering
     \begin{subfigure}[b]{0.45\textwidth}
         \centering
         \includegraphics[width=\textwidth]{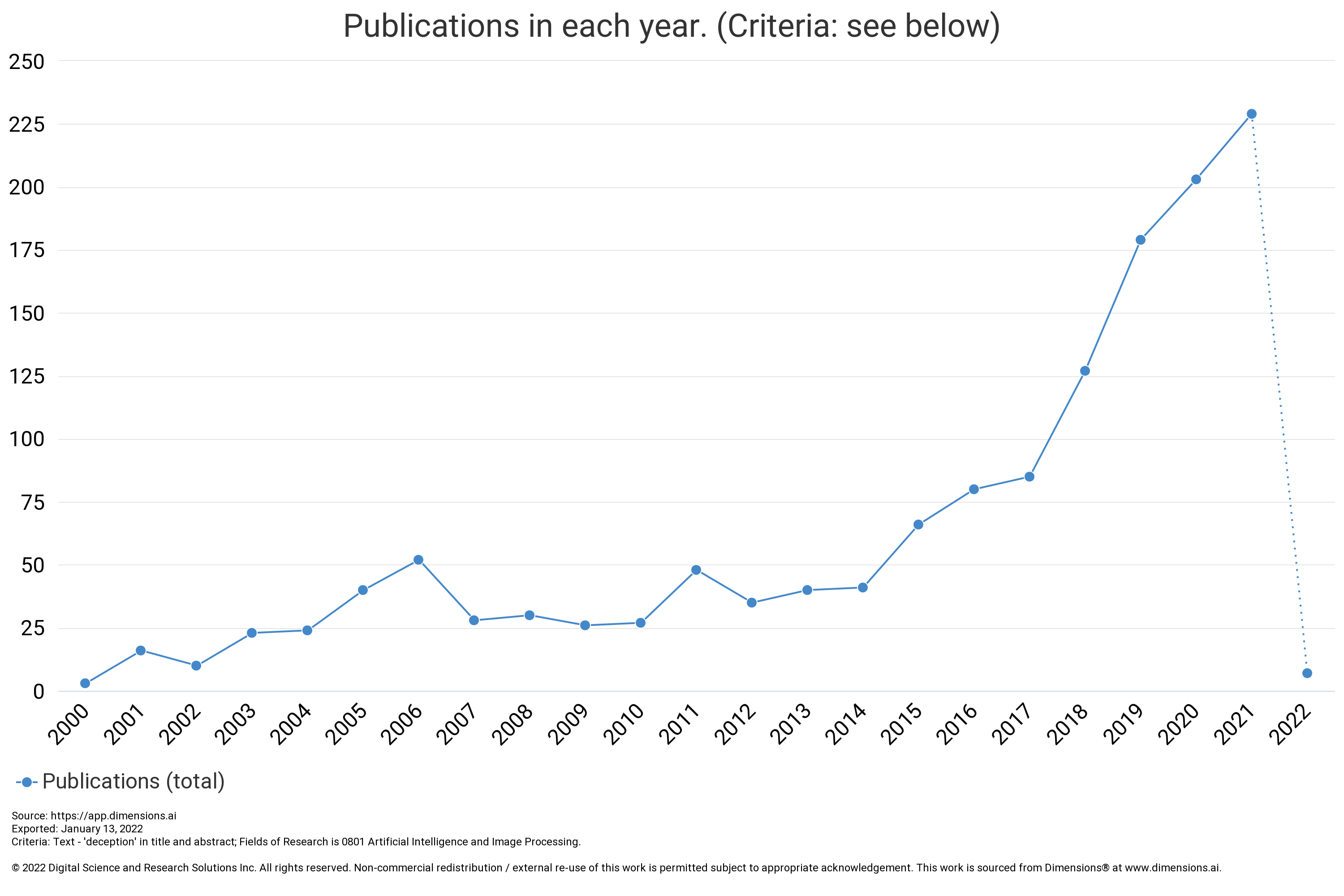}
    \caption{Total number of publications (\textbf{1480}) including the term \textbf{`deception'}.}
    \label{fig:pubs_deception}
     \end{subfigure}
     \begin{subfigure}[b]{0.45\textwidth}
         \centering
         \includegraphics[width=\textwidth]{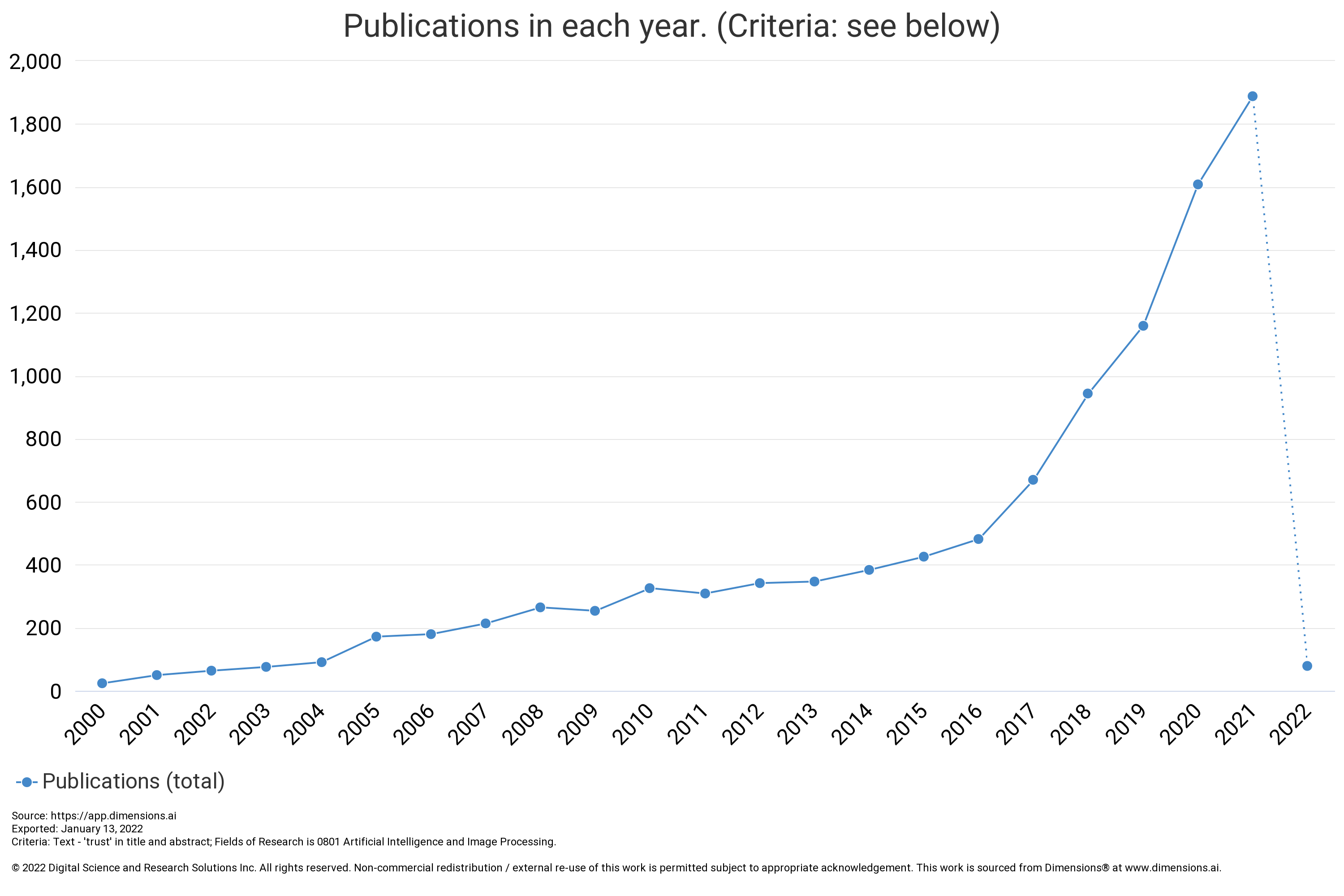}
    \caption{Total number of publications (\textbf{10493}) including the term \textbf{`trust'}.}
    \label{fig:pubs_trust}
     \end{subfigure}
     \caption{Number of publications with respective terms in the title and abstract categorised in the field of AI from 2000 until 2022. The search engine for publications used is www.dimensions.ai, the area of research used to filter results was `0801 Artificial Intelligence and Image Processing'. Chart was printed on 13th January 2022.}
\end{figure}

Research in multi-agent systems (MAS) aims to build models that integrate the social, behavioural and cognitive components of trust and deception. About 20 years ago, \citep{castelfranchi2001trust,castelfranchi2002role} emphasised that this is most important in hybrid human-agent interactions, raising the issue that in order for agents (human or artificial) to reason about the trustworthiness of their counterparts in different contexts, a theory of both trust and deception is necessary. \citep{castelfranchi2002role} indicated multiple levels of trust such as trust in one's agent and mediating agents, trust in the MAS environment and infrastructure, trust in potential partner (collaborator) agents, and finally trust in authorities. Their perspective is reinforced by the one presented by \citep{falcone2001trust} on trust in cyber-societies, which is becoming more relevant given the increasing number and complexity of hybrid interactions between humans and artificial agents. So far, most work has focused on modelling various aspects of trust, independent of deception, as can be observed by comparing the trend in Fig.\ref{fig:pubs_deception} to the one in Fig.\ref{fig:pubs_trust}. Hence, the development of deceptive machines for the purpose of building a holistic socio-cognitive theory has been scarce, and when it did happen, the focus was mostly on particular and very narrow aspects of deception. 

Why is this a problem? Because due to its emergence in hybrid societies, deception can lead to a Tragedy of The Digital Commons, where AI agents maliciously exploit and pollute information shared with humans and other AI agents \citep{greco2004tragedy,sarkadi2021evolution}.  If we want to understand its conditions, consequences, when, why, how, to whom it happens and who is responsible for it, then most of deception's components and forms cannot be treated independently.

In the real world, the area of Intelligence Analysis tries to understand complex phenomena holistically - in various scenarios and contexts, and weighs the risks and impacts of these phenomena. One of the most difficult of these phenomena to explain is considered to be the one of deception. Hence, to understand it holistically, the Intelligence community has developed rigorous methodologies for performing the process of \textbf{inference to the best explanation} (IBE) \citep{Heuer1999psychology}. But how do analysts actually explain and make sense of complex and potentially deceptive agent interactions? According to \citep{ferris1989intelligence}, intelligence analysts and historians of deception try to understand two processes, namely what \textbf{causes} deception and what \textbf{prevents} deception from being caused. Causation and causation prevention of deception also need to be understood in context and as part of an overarching narrative. The main methods for understanding deception in Intelligence Analysis come in the form of structured analytic techniques (SATs - not to be confused with SAT solvers in Computer Science!), most notably the Analysis of Competing Hypotheses (ACH) and its derivations that are used by intelligence analysts to detect deception and reduce cognitive load and bias \citep{Heuer1999psychology}. However, despite being designed to exhaust all possible scenarios of deception and deception causation prevention in complex cases, these methods have proved to be rather tedious to perform by analysts - they are cognitively demanding \citep{vangeld2008can,pope2006formal}.

In this paper, we aim to define an AI-based framework, namely DAMAS, which could significantly enhance the modelling and explanation of deception in hybrid societies, by following the perspective of Intelligence Analysis.

  \begin{figure}
     \centering
     \includegraphics[height=8cm, width=\linewidth]{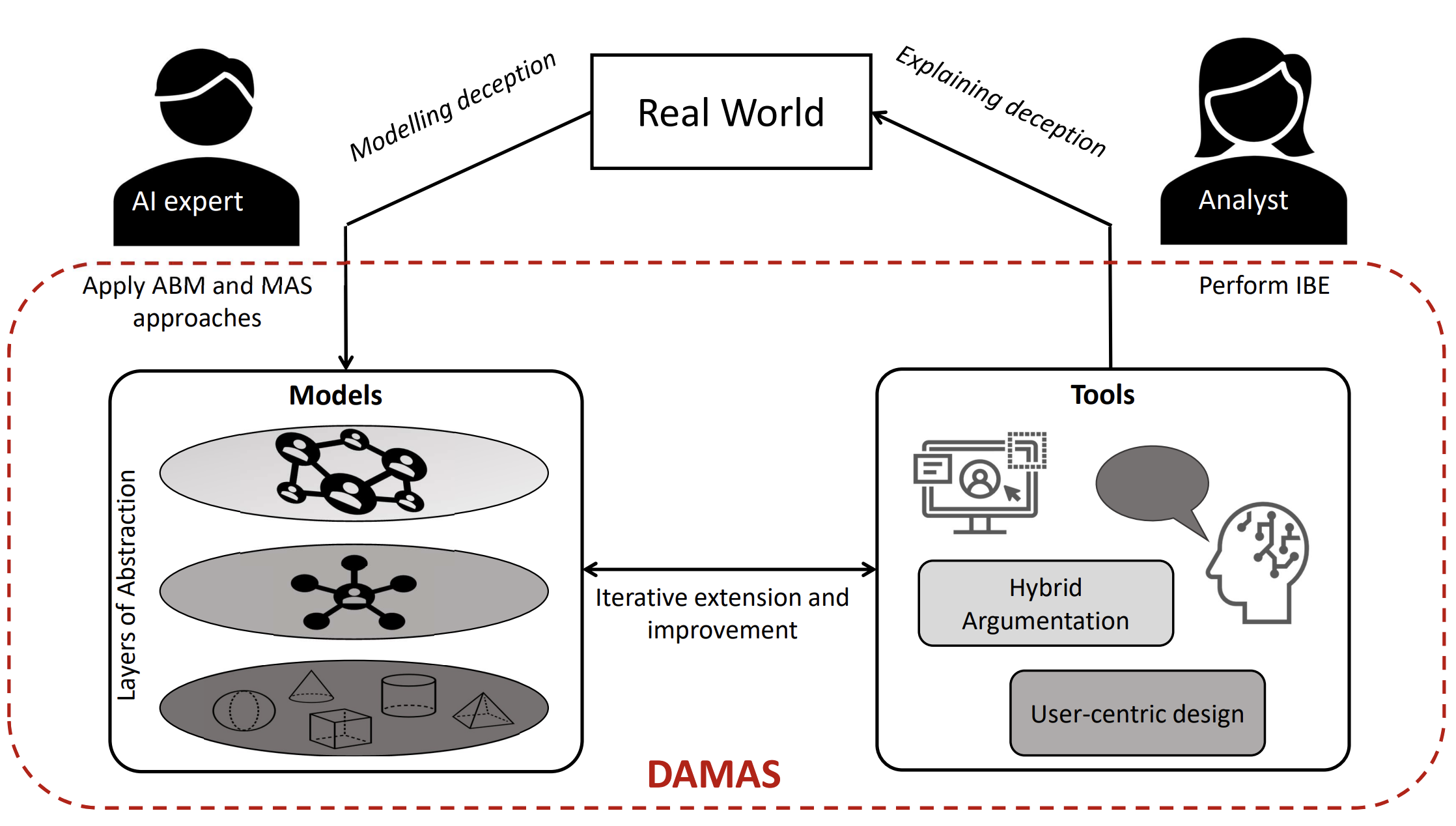}
     \caption{DAMAS would act as a bridge between AI expertise and the methods used in intelligence analysis to explain deception holistically using IBE. }
     \label{fig:narrdamas}
 \end{figure}

\section{Theories and Components of Deception} \label{sec:theories}

To be able to model a phenomenon within a system, we must first understand what that phenomenon means, e.g. have a clear definition of the phenomenon and its eventual underlying components. In this section we will introduce the reader to the main philosophical definitions and psychological approaches in the study of deception. These are necessary for introducing the anthropological view on deception, namely how deception is understood in relation to human nature. What we also introduce are various terms and components that are summoned to describe deception, and we will show how these terms mostly refer to socio-cognitive processes in humans. These terms and definitions are used to refer to various forms of deceptive AI in hybrid societies and understand deception from a computational socio-cognitive perspective, crucial for proposing the DAMAS framework.

\subsection{Philosophical Perspectives} \label{sec:philo}

Philosophy has mostly dealt with definitions of deception. What is a lie? What is bullshit? Are they different? What makes them different? How do lies and bullshit differ from deception? Is deception intentional, e.g. must there be agency and intent behind it? Or can deception happen without intention?

The following, epistemically exhaustive definition of deception is provided by \citep{chisholm1977intent}, and  by  \citep{sep-lying-definition}:

\begin{quote}
    ``[...]to cause another person to acquire a false belief, or to continue to have a false belief, or to cease to have a true belief, or be prevented from acquiring a true belief, or to allow another person to acquire a false belief, or to continue to have a false belief, or to cease to have a true belief, or be prevented from acquiring a true belief."
\end{quote}

If we are careful, we can observe that the definition above merely enumerates the modalities of deception. That is, it describes a category of dishonest communicative behaviours, e.g. to palter, to pander, to lie by omission, to lie by commission, telling half-truths, etc. that can be used to deceive. While an enumeration or a taxonomy of all the ways in which a phenomenon may be instantiated can be useful to understand the phenomenon of deception, according to some researchers in Communication Theory it is by no means a good or proper definition of the phenomenon in itself~\citep{levine2019duped}. What\citep{levine2019duped} suggests is that a functional definition of deception would be more appropriate than merely laying down a behavioural taxonomy. A functional perspective on deception is also endorsed in Philosophy by \citep{artiga2018deception}. Taking their suggestions into account, a functional definition that we adopt to define deception and to refer to deception throughout the paper, without reducing the scope of the discussion to specific types of communicative social behaviour that might be used for deception, is the following:

\emph{The intentional process of an agent, the Deceiver, to make another agent, the Target, to believe something is true (false) that the Deceiver believes is false (true), with the aim of achieving an ulterior goal or desire.}

To summarise, philosophical approaches of deception tend to discuss how various components, namely linguistic, epistemic, ontological, pragmatic etc. determine the adoption of different definitions of deception. So far, the most persistent component has been the intent to deceive (to cause a false belief in the target's mind). Based on the intentional component of deception, functional definitions can be adopted as a general definition of deception to be used in AI.

\subsection{Socio-Cognitive Perspectives} \label{sec:psych}

We now turn to the Socio-Cognitive perspectives on deception, which are mainly derived from theories in Social Psychology and Communication Theory. These areas of research inform us how to understand human deception, how to detect it, and describe what cognitive mechanisms are responsible for deceptive behaviour and communication. There are two main paradigms in the psychology of deception, namely the \textit{cue-based} approaches, and \textit{non cue-based} approaches.

The aim of this paper is to argue for the building of DAMAS, which aims to enhance deception analysis. In order to introduce the socio-cognitive foundations of DAMAS, we will mostly discuss (a) the non-cue based theories of deception and (b) the ability of humans to model the minds of others. The reason for prioritising non-cue based approaches over cue-based ones is that the latter have a significant drawback regarding deception analysis - they are bias prone. Cue-based approaches mostly try to treat passively observable information that represents what lies on the surface of a much more complex socio-cognitive process. Cues can be either verbal - what a person says (linguistic behaviour) or non-verbal - the non-linguistic behaviour of a person, e.g. micro-expressions.  This type of approach solely considers behaviour that is expressed by human agents in order to establish whether the human is being deceptive or not, but it fails to take into account everything else, e.g. the knowledge involved as well as the context of the situations.

This simplistic view of deception, especially the over-reliance on non-verbal cues as micro-expressions, had been extremely popular for some time due to the work of \citep{Ekman1969}. Today, however, there is relatively strong consensus in human deception research that cue-based approaches are very limited and can help detect deception only slightly better than chance \citep{levine2015new}. Notably, the work of \citep{depaulo1996lying,depaulo2003cues,bond2008individual}, has shed some light on how the over-reliance on specific cues is problematic and that these do not improve deception detection. This makes cue-based deception research highly susceptible to cognitive biases such as the truth bias or the confirmation bias (a.k.a. the Othello Error) \citep{bond1987false}.

Alternatively, non-cue based approaches, instead of trying to solely focus on deception detection, they aim to enrich the understanding of various factors and components that underlie social interaction. The following non-cue based theories are actually the ones that can provide with a truly socio-cognitive understanding of deception: \textit{Information Manipulation Theory 2} (IMT2) proposed by \citep{mccornack2014information} and \textit{Truth-Default Theory} (TDT) proposed by \citep{Levine2014truthdefault}. IMT2 aims to understand the cognitive processes that are responsible with speech production in deceptive interactions. One of the arguments proposed by IMT2 is that there is no difference w.r.t. cognitive load when humans lie or tell the truth. TDT differs from IMT2, by focusing on the contextual aspects of deception. TDT proposes that there is a general \textit{truth-default} state in which humans are susceptible to deception and from which they can be triggered out of depending on the informational context. In addition to IMT2 and TDT, there is \textit{Interpersonal Deception Theory} (IDT) proposed by \citep{Burgoon1996deceptive}, which can provide us with a rich socio-cognitive perspective. While IDT is cue-based to some extent, it is not over-reliant on cues. This theory is very different from standard cue-based approaches and its strength is that it aims to give an integrated perspective on how cues influence inter-personal interactions between human agents. In one way, IDT represents deceptive interactions as a multi-agent system that is governed by rules of message transfer.

There is a common component throughout all of the non-cue based perspectives on deception, namely the ability to form, simulate, and reason about the minds of others, which is known in the scientific literature as the ability to use Theory-of-Mind (ToM). Different `flavours' of ToM have been proposed to explain this phenomenon. One flavour is \textit{Theory-Theory of Mind} (TT). TT represents the type of ToM where agents already have an understanding to some extent of the other agents' minds. TT can be, for instance, a set or a knowledge base that consists of beliefs of others' mental attitudes. Another flavour is known as \textit{Simulation Theory of Mind} (ST). ST represents a process for using the already known set of others' mental attitudes to simulate their minds in order to see what other attitudes they can form or to predict their behaviour. Both types of ToM have emerged from the study of children in developmental psychology \citep{gopnik1992child}. \citeauthor{gopnik1992child} describe how the two competing flavours aim to explain how children predict others' behaviour, e.g. whether they use language associated to TT such as `beliefs' and `desires' - theoretical language constructs that form a conceptual representation of another's mind, or whether they use phenomenal constructs associated to ST by using their own minds to `run' a simulation of inputs and generate outputs in the form of an `experience'. In TT, the child heavily relies on a conceptual constructs and language, whereas in ST egocentrism plays a crucial role, as the child heavily relies on projecting one's own mental states onto another.

Even if there has been an ongoing debate namely between the proponents of TT and the proponents of ST regarding how humans develop and use ToM, these perspectives seem to be merging into a hybrid perspective on ToM \citep{Goldman2012theory}. While TT represents the type of ToM where agents already have an understanding to some extent of the other agents' minds, e.g., a set or a knowledge base that consists of beliefs of others' mental attitudes, ST represents a process for using the already known set of others' mental attitudes to simulate their minds in order to see what other attitudes they can form or to predict their behaviour. TT and ST can be merged into hybrid ToM, which can be thought of as a `high-level' ToM in the form of a simulation setup in which an agent takes its target's beliefs, desires, and intentions as premises for generating a practical reasoning process in the way its target (the other agent who's mind is being mentalised) would \citep{goldman2006simulating,Goldman2012theory}.

Some of the other socio-cognitive components of deception have been described by IMT2 to consider how information or knowledge is managed inside the deceivers' cognitive apparatus. This cognitive apparatus in humans represents some sort of cognitive agent architecture that is able to run various processes before, after, or during communicative interactions. IMT2 describes it as `\textit{a speech production system 
involving parallel-distributed-processing guided by efficiency, memory, and means-ends reasoning; and this production process involves a rapid-fire series of cognitive cycles (involving distinct modules united by a conscious work-space), and modification of incrementally-constructed discourse during the turn-at-talk in response to dynamic current-state/end-state discrepancies}'. IMT2 identifies two main cognitive processes derived from speech-act theory of cognition that are responsible for information selection and dissemination. The first process is called \textit{Pars Pro Toto}, meaning the parts for the whole, which a sender agent uses to select the information that, given the context in which the sender and the receiver find themselves, will convey the sender's intended meaning of the message in the mind of the receiver. The second process is called \textit{Totum Ex Parte}, meaning the whole from the parts, which a receiver uses to infer the intended meaning of the message from a sender given the context in which the receiver and the sender find themselves in. According to IMT2, deceivers and deception detectors engage in these two processes when they interact. Fascinatingly, IMT2 has been inspired by work at the intersection of AI and pragmatics \citep{newell1972human,hovy1990pragmatics}, hence its adoption of specialised terms from AI when describing the speech production systems for humans. Unsurprisingly to people working in AI, IMT2 seems highly tuned to the idea of intelligent socially-aware AI agents \citep{cohen1985speech}.  

On the other hand, IDT deals with deception at a different level of abstraction from IMT2, as it ignores the actual information that is managed by the internal cognitive processes of humans, and focuses instead on the socio-cognitive factors that constrain speech production. For starters, \textit{cognitive load} is a crucial factor in deceptive interactions that determines the success or failure of the deceiver. From a computational perspective, cognitive load represents the number and complexity of operations an agent needs to perform on a certain quantity of information in order to deceive or detect deception.
According to IDT, the way in which humans are able to cope with cognitive load varies between individuals. IDT identifies \textit{communicative skill} of the agents as the most important factor that influences this ability. Agents with a high communicative skill tend to be better at managing the cognitive load. Computationally, the communicative skill might be represented either by some amount of computational resources that are available to an agent, or by some communication mechanisms that are more or less efficient (probably depending on the circumstances/contexts) in disseminating information and that are part of an agent's cognitive architecture. Another important parameter is the amount of \textit{leakage}, that represents cues that contradict a deceptive message, and that an agent exhibits during interactions. IDT argues that leakage increases with cognitive load, but decreases with communicative skill. Agents with high communicative skill, being able to manage their cognitive load, can reduce the amount of leakage.

Furthermore, TDT \citep{levine2019duped} describes the contextual level of deception, which informs us that there is more to deceptive interactions than social parameters like cognitive load and leakage when it comes to interpreting the information received or transmitted. According to TDT, what is communicated (informational content) in a given context should be given more importance than non-verbal behaviour that may or may not happen to be correlated with the communicated content. Therefore, based on TDT, the contextual knowledge and information available to the agents that interact with each other should be a crucial component of computational deception. The way this information is used can determine whether deceptive attempts are successful or not. The same goes for attempts at detecting deception, such as persuading an agent to reveal deceptive motives. This information should represent knowledge about what is said or not, when and where it is said (or not), to whom (and not), as well as how it is or not interpreted by someone. This type of information can trigger agents into or out of the \textit{truth-default} mental state, which makes individuals spend less or more cognitive resources on what is being communicated, e.g. makes them more or less suspicious to others depending on the context.

Even though IMT2 and IDT do not mention ToM as an explicit component of deception, ToM seems to be assumed implicitly. For instance, as a `high-level simulation theory', hybrid ToM blends TT and ST into practical reasoning \citep{goldman2006simulating,Goldman2012theory}, which is a process very similar, if not identical, to the processes of \textit{Pars Pro Toto} and \textit{Totum Ex Parte} described by IMT2. Now, the only theory of deception out of the three described here, that explicitly mentions ToM is TDT \citep{levine2019duped}, where deception would be a special case of ToM. This seems to be a well-founded consideration given that humans develop the ability to form and use ToM in early childhood. Another strong argument for the tight link between human ToM and deception is the use of false-belief experiments to track the formation of ToM in children \citep{gopnik2012reconstructing}. Based on this, we can look at the definition of deception (see Section \ref{sec:philo}) and safely say that intentional exploitation of false belief formation is nothing other than deception, confirming Levine's point that deception is a particular case of ToM in use.

To summarise, there are two main approaches to understanding deception in Psychology, namely \textit{cue-based} and \textit{non cue-based}. While most cue-based approaches suffer from numerous drawbacks, including heavy biases in deception detection, non cue-based approaches have been designed to overcome these and explain deceptive interactions more rigorously. Hence, these are the approaches that need to be facilitated by DAMAS in order to support the analysis of deception in hybrid societies. Moreover, there is the human ability of using ToM which plays a crucial role in deception. In particular, the emerging hybrid flavour of ToM seems to be most appropriate to be represented with AI approaches within DAMAS, especially regarding the practical reasoning processes involved in agent-to-agent deceptive communication.

\section{Reductionism \& Previous Research in Deceptive AI}\label{subsec:AIres}

The literature in Social Psychology and Communication Theory proves to be critical in understanding deception as a socio-cognitive interactive process between agents, starting from the internal cognitive mechanisms responsible for the production of communicative behaviour, to giving an overarching perspective between the relations of the agents involved in deceptive interactions. In the previous section, we have seen that there are multiple socio-cognitive components of deception in the literature that are relevant. The most crucial of these components are ToM, the social factors  and the idea of agent-agent interpersonal dynamics described in IDT, the reasoning processes for speech production described by IMT2, and the importance of contextual factors and truth bias emphasised by TDT. If we are to build a holistic AI framework to study deception in hybrid societies, we must design it such that it is able to represent and model these components as part of multi-agent interactions and cognitive agent architectures. Then, what we need to do with AI methods is to capture and understand the relation between the socio-cognitive components of deception and ToM inside hybrid societies.

\subsection{Multi-Agent Frameworks vs AI Reductionism}

There exists a methodological, or better said a paradigmatic limitation to building holistic frameworks for modelling deception in AI. This limitation comes from the fact that AI approaches are usually highly reductionistic, namely that real world factors are abstracted away for the purpose of ``elegance'' and simplicity of AI models. In the history of science and philosophy, reductionism has been extensively debated for hundreds of years \citep{andersen2001history}. In modern science, \citep{marr2010vision} started discussing the limitations of reducing cognitive functions to physical properties of neurons, arguing that explanations about cognitive functions cannot solely rely on neuro-chemical activity \citep{peebles2015thirty,bickle2015marr}, but that in order to be meaningfully understood they also require references to behaviour as part of an overarching multi-level analysis \citep{marr2010vision,krakauer2017neuroscience}.

A similar argument can be made about research in AI. That is if we want to understand and explain a socio-cognitive phenomenon such as deception, for which other areas of scientific and humanities research have identified various components that interact with each other to give rise to a complex multi-layered process inside complex multi-layered systems that are hybrid societies, then we cannot reduce these types of processes to a single function. Historically, AI has been all about breaking down human intelligence and modelling it computationally as functions at a very abstract level \citep{boden1996artificial}. This is understandable for investigating the properties of some processes in isolation. But when a process consists of a multitude of these functions working sequentially and in parallel at multiple levels of abstraction - from the internal cognitive reasoning mechanisms to the interaction mechanisms responsible for communication to the emergent large-scale behaviour of agents that are interacting - then just solely focusing on the properties of a single computational function, independently of other components, seems to miss the point of modelling the entire phenomenon itself\footnote{The same argument applies to why definitions that explain the entire phenomenon of deception are better suited than others which only define a particular type of deceptive behaviour.}. Aside from representing the functions themselves, there is the problem of a causality between elements at various layers of abstraction, e.g. the problem of how functions relate to each other. In turn, this means that AI models need to account for different kinds of architectures. This is why the area of Distributed AI, and later ABM and MAS, emerged to be able to study systems of agents. However, even in these areas, most of the work has lost focus of socio-cognitive modelling and has turned to benchmarking and optimisation of groups of artificial agents \citep{dignum2020agents}.

Despite the limitations above, Multi-Agent Systems (MAS) frameworks have numerous benefits, of which most crucial for understanding complex interactions is their ability to capture these interactions at different layers and levels of abstraction. MAS and agent-based modelling (ABM) techniques are being increasingly applied in the real-world in finance, politics, traffic control, industrial development, urban planning, military operations, cybersecurity etc. Why not develop a framework that can be applied in deception analysis? If appropriately developed, DAMAS could be used to automate some, if not most, of the tedious tasks performed by deception analysts and ensure bias-free intelligence analysis of deception. The main benefit of having a framework to critically analyse deception is that it can be used for high-level decision making, especially in the domain of regulating AI and preventing risks coming from deceptive AI agents.

One method in MAS to limit the negative effects of reductionism is to model and integrate \textbf{layers of abstraction}, which are the foundations of a MAS framework \citep{JaCaMo-2013}. Their integration accounts for a rich multi-dimensional world instantiated through computational representations of environments, agents, and organisations. Social interactions between agents are complex and some of the properties of these interactions and the relations between these properties can only be described by taking into account these different layers. As an illustrative example from MAS engineering, the JaCaMo framework ~\citep{JaCaMo-2013} integrates the three layers of abstraction respectively through the Jason agent oriented programming language (agent layer), the CArtAgO common artifact infrastructure for agents open environments (environment layer), and the Moise organisation oriented programming framework (organisation layer). Various MAS models can be then represented and implemented using such a framework to explore agent systems at the level of interaction, considering how agents interact with the environment and each other as part of various organisations that follow different norms etc.

Another way to deal with reductionism in AI is to account for multiple \textbf{models and levels of abstraction} through a combination of MAS and ABM approaches, representations, and techniques, formal or computational, that can be used to describe the properties of interactions between agents. MAS approaches focus on modelling the \textbf{interaction level}, while ABM approaches focus on modelling the \textbf{descriptive level} \citep{dignum2020agents}. An example of a MAS model that focuses on interaction between agents is a computational model based on speech-acts. Such a model describes and/or predicts both the cognitive processes and the behaviour used to produce changes in the belief bases of other agents through speech-production. This type of model, for instance, can be used to track the information exchanged during a dialogue, as well as the reasoning that the agents perform during the dialogue. On the other hand, an ABM model that focuses on describing emergent behaviour does not need to represent every cognitive component responsible for producing a speech-act, it only needs to describe an overall state of a system. Thus, it can be efficiently used to represent the costs and benefits of communication and compute states of the system at different moments through software simulation. For instance, it can check if an equilibrium is ever reached for exchanging information given a set of social parameters.

Therefore, if we want to limit scientific reductionism when analysing deception, then it is necessary to model deception holistically, as part of an over-arching socio-cognitive process taking place inside a multi-agent system that integrates multiple layers of abstraction. This is not only valid for human-to-human interactions, but also for machine-to-human, machine-to-machine, or any type of agent-to-agent interactions. That is why we propose the development of DAMAS.

\subsection{Previous Work}

In the last 20 years, the importance of research on deceptive AI has been emphasised in the AI community during a series of events, namely the 2015 AAAI Fall Symposium on Deceptive and Counter-Deceptive Machines\footnote{See the proceedings at https://aaai.org/Press/Reports/Symposia/Fall/fs-15-03.php.}, and the two International Workshops on Deceptive AI\footnote{See the joint proceedings in \citep{sarkadi2021deceptivebook}.} co-located with ECAI2020 and IJCAI2021. Having gone through the socio-cognitive components that need to be considered when modelling deceptive AI, we now describe the work at the intersection of deception and AI, grouping them into six categories to better understand what computational aspects and properties deception they address.

With some exceptions, most of the approaches presented in this section focus on narrow aspects of deception and they are highly reductionistic. Where one approach manages to capture a certain aspect, it fails to consider a multitude of other aspects and components of deception. However, these aspects and properties also need to be understood to inform the development of purpose-specific models or tools for the holistic socio-cognitive framework (DAMAS). The purpose of this section is not to criticise previous work, but to explain why it is reductionistic and how it could benefit from being integrated under an agent-based conceptual framework, namely DAMAS.

There is some work at the intersection of deception detection and Machine Learning that we left out in this section due to its over-reliance on cue-based theories of deception. The work that we do summarise covers various aspects of deception in AI going beyond the literature presented during the three events mentioned above (for a quick overview of the covered literature, the reader can consult Table \ref{tab:airesearch}):

\subsubsection{Deception in Agent Societies} 

Research that falls in this area aims to motivate the AI community to build a theory of trust and deception in virtual societies. The main body of research mainly focuses on human-computer interaction in MAS and tries to answer questions that arise from interactions between human and artificial agents, such as \textit{Will artificial agents deceive?}, \textit{Why would they deceive?}, \textit{Why is it important for agents to be able to reason about deception?}, \textit{Why should we model deceptive agents?} \textit{What are the types of deception in virtual communities?} \textit{What is the role of deception in human-agent interaction?}

Recent findings back up this perspective by showing how humans and artificial agents influence each other's attitudes and behaviour. For example, \citep{mell2018welcome} focus on how the negotiation strategies of artificial agents determine humans to endorse artificial deception. Apparently and counter-intuitively, humans also tend to be more cooperative with deceptive machines than with friendly ones, according to \citep{ishowo2019behavioural}. Another type of study, presented by \citep{dras2010generating}, looks at the ways in which deceptive language can be generated and detected in virtual systems. 
    
These approaches can inform us about how humans can distinguish between deceptive and non-deceptive AI agents in dialogues; that deceptive agents are better at inducing cooperation; and the more experienced human negotiators become, the more likely they are to endorse deception performed by their AI agent representatives. However, the downside of these methods is that they study deception as part of isolated interactions, despite being aimed at studying deception as part of complex hybrid society. On the other hand, these approaches can inform us on the modelling of human-machine relations within DAMAS.

\subsubsection{Logical Aspects of Agent Deception} 

This area aims to define a logical taxonomy of deceptive behaviour, as well as representing and modelling deception using logical formalisations. Some relevant questions are: \textit{What is the the difference between lying and deceiving?}, \textit{How do we represent bullshit, pandering, paltering etc.?}
    
Works in the areas of Logic and Philosophy have strongly influenced the current state of the art approaches in deceptive AI. Most known are the works in knowledge representation, which have laid the foundations to formalise deceptive communication. These appear in \citep{sakama2010logical}, in \citep{sakama2015formal}, and in \citep{sakama2015formalIGPL}, where the authors present the logical distinctions between different types of dishonest linguistic behaviour, and later introducing a model of causality for them \citep{sakama2021};\citep{bonnet2021} sketch a logical theory of belief manipulation; \citep{van2014dynamics} introduce a dynamic logic of lying that considers lies about factual propositions as well as lies about the beliefs of others; in \citep{uckelman2011deceit} where the author uses the medieval concept of \textit{dubitatio} to study deceptive agents; \citep{Jones2015} proposes a formal model of self-deception that is consistent given certain epistemic conditions imposed on the agent that engages in self-deception; and \citep{smith2016construction} describe a logic-based account of analysing conjuring tricks as form of deception, and identify substitutable elements and stable occlusion as crucial components to construct perceived impossibilities.
    
Logic-based accounts of dishonesty enable us to elegantly and rigorously capture communicative accounts of deception. However, these formalisms lack the ability to model or represent socio-cognitive aspects such as degrees of trust, bias, and ToM. The trade-off for using such approaches is between representational power and rigour. On the other hand, because these approaches can inform us how to rigorously formalise communicative acts and arguments, they could be used to model deceptive argumentation between agents in DAMAS.
    
\subsubsection{Strategies for Agent Deception} This area studies deceptive strategies in MAS. This area is strongly influenced by Economics and Cybersecurity, from which methods such as risk and threat modelling have been adopted and adapted to address specific scenarios. Some relevant questions are: \textit{What are the deceptive strategies and counter-strategies in different contexts?}, \textit{How to reduce and mitigate deceptive attacks?}, \textit{How do we design a system that either reduces or incentivises agent deception?} etc.
    
Some of the work in this area explores issues such as using heuristics to cause target agents to execute a plan that will achieve the deceiver's desired goal \citep{christian2004strategic}; finding deceptive strategies using path-planning \citep{masters2017deceptive,masters2021extended,masplan2021}; using a MAS system on a Bayesian network test-bed to distinguish between truthful and deceptive agents based on the correlation of the agents' beliefs \citep{santos2009deception}; modelling agent based deceptive interactions on social networks \citep{barrio2015dynamics}; modelling deceptive interactions to counter reconnaissance-based cyber-attacks \citep{schlenker2018deceiving}; studying the effects of deception in repeated games between learning agents \citep{nguyen2019deep}; applying IDT in multi-agent reinforcement learning to reduce the effects of social engineering attacks \citep{yang2018use}; developing an information-theoretic model of deceptive strategies \citep{kopp2018information}; learning deceptive strategies through bayesian belief manipulation \citep{learningdec2021}; formalising cyber-deception games between multiple agents using hyper-game theory methods \citep{ferguson2019game}. Additionally, given the increasing interest in explainability for AI,  \citep{wright2019agents} discuss how agents can strategically provide users with deceptive and rebellious explanations.
    
These approaches allow us to discover optimal strategies for deceptive behaviour of a single agent and find equilibria when more than one agent is involved. Usually these methods focus on some utility-based (cost and reward) metric to study strategies. The downside of these approaches is that they focus more on descriptive approaches rather than interactional - there is no representation, architecture or model of the agents' knowledge that changes - these are just implicitly assumed in the modelling of strategies. Hence, in terms of trade-off, these approaches sacrifice representation and generality for simplicity and efficiency in specific and well-defined setups. W.r.t. DAMAS, these approaches can help us identify scenarios where agents make fully rational (from a utilitarian viewpoint) decisions and operationalise rational agent behaviour.

\subsubsection{Reasoning in Agent Deception}
This area looks at formal and informal reasoning mechanisms responsible for deception, and is strongly influenced by Informal Logic and Argumentation as well as by subareas of Cognitive Science. Some relevant questions are: \textit{What are the cognitive components involved in deceptive reasoning?}, \textit{What types of reasoning mechanisms are involved in deceptive interactions?}, \textit{What type of cognitive architectures can be used to represent deceptive reasoning?}, \textit{What type of knowledge is necessary or not for an agent to deceive?} etc.
    
Work in this area of deceptive AI has explored issues such as using and detecting deception in argument debate games by \citep{sakama2012dishonest} that are formalised using abstract argumentation \citep{Dung-1995-OAAFRNRLPNG}; using abductive reasoning for deception \citep{sakama2011dishonest}; using argument mining for detecting deceptive reviews online \citep{cocarascu2016detecting}; using argument mining for detecting propaganda \citep{vorakitphan2021don}; using argumentation-based tools to analyse disinformation in fake news \citep{delobelle2020sifting}; looking into what type of arguments can be used by a machine to deceive \citep{clark2010cognitive}; modelling agents that use mindreading for deception \citep{isaac2014mindreading}; modelling the meta-reasoning of agents that tell deceptive stories and detect deceptive stories in dialogue argumentation games \citep{sarkadi2019DecStoryAAAI}. 

These methods enable us to (i) capture informal aspects of reasoning involved in both deception and deception detection, (ii) represent the cognitive processes such as meta-reasoning, and (iii) use automated techniques to categorise deceptive communication. Argument mining techniques are used to detect potentially deceptive arguments from text, whereas knowledge representation and reasoning techniques are used to model the internal reasoning mechanisms responsible for producing deceptive communication in agents. The benefits of these approaches is that they can capture the links between deceptive communication, its detection, and their respective reasoning mechanisms and cognitive biases. The downside in terms of reductionism is that they are not integrated as part of an overarching system. On the other hand, these approaches can help us represent and operationalise complex forms of reasoning and decision-making and identify instances of where irrationality is involved (irrational as non-utility oriented) within DAMAS.

\subsubsection{Engineering Deceptive Agents} 
This area looks into the design, modelling, and engineering methods that can be used to create deceptive or deception-detective autonomous agents. The main difference between the area of engineering of deceptive agents and the areas covering strategies and reasoning of agents is that the engineering approach aims to integrate these agents into more complex systems. This area does not just look at how agents reason about deception, but what causes them to achieve deception or not in a given system. Some relevant questions are: \textit{What are the best approaches to model deceptive agents or deceptive interactions without sacrificing expressivity?}, \textit{How should these models or systems be designed and implemented?}, \textit{How do we evaluate these systems?}, \textit{What can we use these systems for?}, and \textit{How do we integrate different components of deceptive interactions in order to engineer complex reasoning agents?} etc.
    
Recent work in this area has looked at issues such as: modelling agent architectures that engage in deceptive storytelling during dialogues \citep{rato2017strategically}; developing deceptive chatbots for emotional well-being and mental health \citep{efriend2021}; engineering agents that use lies, bullshit and deception \citep{PanissonSarkadi-2018-LBandDinAOPL}; and also the first attempt in MAS to integrate IMT2, TDT, and IDT in order to model, implement and evaluate deceptive interactions between complex reasoning agents that use \textit{Theory of Mind} by \citep{sarkadi2019dectom}. 
    
These methods enable us to specify models, design agents, and implement them as prototypes to study deceptive interactions between them. The benefits of these approaches are that they are modular, engineering-driven, and based on reasoning mechanisms. Most benefits can be seen in the approaches that try to integrate socio-cognitive models in their multi-agent models which are general enough to be applied to a multitude of contexts, whereas most downsides w.r.t. reductionism come from the fact that despite the generality, interpretability, and applicability of the models, their domain-specific implementations usually lack automated and user-friendly explanations of the agent interactions. On the other hand, these approaches can be used as the building blocks for modelling multi-layered cognitive processes, architectures, and communicative behaviour of agents within DAMAS.

\subsubsection{Embodied Deceptive Agents} This area studies the performance of deceptive robot agents in the physical world. Some relevant questions are: \textit{How can robots deceive humans?}, \textit{What strategies can a robot employ for deception?}, \textit{What kind of robots are more deceptive?} etc. This area is becoming increasingly relevant given the advancement of Internet-of-Things and the interconnectedness of physical agents, such as self-driving cars and robot assistants.
    
Research in this area looks at issues such as: how deception about agency influences the way children treat robots on a social level and when is it acceptable to deceive children about the nature of the robots they interact with \citep{westlund2015deceptionkids}; how robots can decide to perform physical actions such that they manipulate the beliefs of humans who observe them \citep{gray2014manipulating}; how to provide robots with the capacity to deceive \citep{wagner2011acting}; why the properties of deception and transparency in robots should be treated in the context of user vulnerability \citep{collins2017vulnerable};  using biologically inspired behaviour for robot deception \citep{shim2012biologically}, such as squirrel behaviour and applying it to shill agents~\citep{arkinshill2021}. \citep{shim2013taxonomy} even provide a taxonomy of robot deception and present the related benefits of deception for human-robot interaction. 
    
These approaches are very useful to study human biases w.r.t. the design and behaviour of robots. Regarding reductionism, the robots are very limited in terms of reasoning and complex communication. This is perfectly understandable since control-based human-robot interactions in labs need to isolate various factors in order to gain significant findings. However, the findings themselves should be used to inform the design of more accurate and complex models of deception in human-agent interactions. For instance, the knowledge that a specific type of embodied AI agent is more likely to mislead, persuade etc. human agents is a socio-cognitive factor and can be used as part of DAMAS to compute the likelihood of deception being caused if that type of agent is involved in some multi-agent interaction.
  
\bigskip
    
We have seen in this section that research on deception involves looking at various aspects and components of deception. However, more often than not, AI does so from a highly reductionist approach. This type of reductionism can be problematic, because compared to other forms of dishonesty such as lying or bullshitting, both of which can be elegantly modelled in logics of communication, deception involves more complex cognitive mechanisms that have to be modelled \citep{PanissonSarkadi-2018-LBandDinAOPL,hyman1990deception}, usually in a relational manner \citep{mchugh2004relational}. For instance, ToM is considered a critical component of deception~\citep{sarkadi2019dectom,sarkadi2020phdthesis,isaac2017,derosis2003can}, and more often than not, models in AI fail to consider ToM as a complex form of socio-cognitive process\footnote{The AI literature has also emphasised the potential benefits of artificial agents with ToM (that is not necessarily high-level simulation), that include the explainability, the efficiency, and the increased social performance of machines. For example, \citep{de2017negotiating} show how agents with higher order ToM (`higher' here means strictly meta-level recursivity over a single belief, not in the sense of `high-level simulation') outperform other agents in negotiation. Another ability is increasing the seamlessness of cooperation between humans and robots \citep{winfield2018experiments,bianco2020psychological}. Such benefits, we believe, might be the reason why considerable efforts are being made in the AI community to enable machines to form and use models of other minds \citep{albrecht2018autonomous}.}. 

In hybrid societies, modelling the ToM component has both risks and benefits, as it allows us to give agents the ability to align their ethical values with humans \citep{isaac2017}, but then it also allows machines to apply higher levels of ToM to outperform humans in deception and counter-deception \citep{sarkadi2020phdthesis}. What we mean here by higher level is either deeper levels of recursivity or a greater number of parallel simulation instances/branches when modelling other minds in the sense of Goldman's `high-level simulation'. As an example, machines could outperform humans using higher-level ToM by exploiting \textit{unknown unknowns} \citep{sarkadi2019dectom}\footnote{`\textit{Reports that say that something hasn't happened are always interesting to me, because as we know, there are known knowns; there are things we know we know. We also know there are known unknowns; that is to say we know there are some things we do not know. But there are also unknown unknowns—the ones we don't know we don't know. And if one looks throughout the history of our country and other free countries, it is the latter category that tends to be the difficult ones.}'\citep{rumsfeld2002unknown}.}. Introduced publicly during a US Dept. of Defense news briefing by Donald Rumsfeld in the context of Intelligence Analysis, the unknown unknown means \textit{not knowing that one does not know something}~\citep{logan2009known,rumsfeld2002unknown}. By exploiting unknown unknowns, AI agents do not simply exploit missing information due to the target's inaccessibility to sources, but also the target's failure to reflect on the information that is or is not available - either due to bias or due to limited cognitive capabilities or physical resources (lack of time or lack of analysis tools). Conclusively, it is unrealistic to model deception in agent societies without having the proper kind of representation and implementation for the deceptive agents, their internal cognitive processes, and their behaviour within the complex system in which they interact.

To summarise, if we are to model deception in hybrid societies from the direction of socio-cognitive multi-agent systems, then a natural step for integrating multiple forms of deception as part of a single over-arching system would be to develop an explainable (analysis-friendly) MAS framework, namely DAMAS. In the next section we will describe the desirable properties of DAMAS and explain how it may account for a multitude of deception components, aspects and agent-oriented models of deception.


\begin{table}[]
    \centering
    \begin{tabular}{|l|p{11cm}|}
    \toprule
    \textbf{Topics} & \textbf{Papers} \\
    \midrule
        Agent Societies & \citep{mell2018welcome,ishowo2019behavioural,dras2010generating,sarkadi2024self} \\
        
        \hline
        
        Logical Aspects &  \citep{sakama2010logical,sakama2015formal,sakama2015formalIGPL,sakama2021,bonnet2021,smith2016construction,van2014dynamics,uckelman2011deceit,Jones2015}\\
        
        \hline
        
        Agent Strategies & \citep{christian2004strategic,masters2017deceptive,learningdec2021,masplan2021,santos2009deception,barrio2015dynamics,schlenker2018deceiving,nguyen2019deep,yang2018use,kopp2018information,ferguson2019game,wright2019agents}\\
        
        \hline
        
        Reasoning & \citep{sakama2012dishonest,Dung-1995-OAAFRNRLPNG,sakama2011dishonest,cocarascu2016detecting,clark2010cognitive,isaac2014mindreading,sarkadi2019DecStoryAAAI}\\
        
        \hline
        
        Engineering & \citep{rato2017strategically,efriend2021,PanissonSarkadi-2018-LBandDinAOPL,sarkadi2019dectom}\\
        
        \hline
        
        Embodiment & \citep{westlund2015deceptionkids,gray2014manipulating,wagner2011acting,shim2012biologically,shim2013taxonomy,arkinshill2021}\\
        
        \hline
        
        Ethics* & \citep{kampik2018coercion,sklar2004okay,isaac2017,collins2017vulnerable,chakraborti2019can,coeckelbergh2018describe,shim2013taxonomy,sharkey2021we,natale2021deceitful,bender2021dangers, sarkadi2023should,sarkadi2023deceptive}\\
        
        \bottomrule
    \end{tabular}
    \caption{The research categories on modelling deception with AI techniques described in section \ref{subsec:AIres}. For a holistic overview of ethics and Deceptive AI, see \citep{sarkadi2023deceptive}.}
    \label{tab:airesearch}
\end{table}


\section{DAMAS: A MAS Framework for Deception Analysis} \label{sec:damas}

When designing a framework that aims to facilitate the analysis of complex phenomena, such as deception, in hybrid societies, it is crucial to keep in mind a number of principles.

First, such a framework should offer enough representational power in order to enable the automation of counterfactual reasoning about social dynamics between agents. This is achieved by representing (i) the environment, (ii) agents and their communicative abilities, (iii) the knowledge that agents have (such as nested beliefs and relations between beliefs and ToM of each other), (iv) how changes in the environment influence the agents' beliefs, and (v) how the agents themselves cause these changes intentionally or unintentionally - all of these in a relational fashion \citep{mchugh2004relational}.

Second, the framework should be both scalable and modular. In other words, it needs to be able to support the seamless addition and integration of new agent, environment, and organisation types. That is, the models of the world should be easily increased and modified based on potential changes in deception modelling requirements. For instance, whenever a new deception model or a new type of deceptive agent architecture is developed, then this should be easily added to the framework without changing the already existing models or agent architectures. The interoperability of agents and tools can guarantee this property. MAS frameworks such as JaCaMo are designed based on these two principles, enabling an almost seamless interoperability between the environment, agent, and organisation layers \citep{JaCaMo-2013}.

Third, such a framework should be usable. Analysts should be able to use the framework to explain why or why not interactions between agents happen the way they happen. Explainable AI (XAI) techniques can be used to extract narratives of agent dynamics and generate explanations in a human-interpretable manner, e.g. by taking into account the principles of causality, counterfactuality, contrastiveness, and social factors to tailor the explanations of agent behaviour based on the user's background knowledge, expertise, and preferences \citep{miller2018explanation}. 

To develop DAMAS is analogous to making the knowledge of AI experts accessible to deception analysts. First, AI experts should create ABM and MAS models to represent the world at multiple socio-cognitive layers of abstraction. Then, these models should be made available to deception analysts through a library of software packages and XAI tools. In this way, analysts would be able to iteratively engage with these tools in order to improve the existing models and adapt them to their needs in order to perform IBE. DAMAS, as a framework, would then act as a mediator or a bridge between the knowledge of AI experts and the expertise of intelligence analysts, which would support the holistic understanding of deception as a socio-cognitive phenomenon in hybrid societies.

By keeping the three design principles in mind, we show an overview of how DAMAS can support the Intelligence Community in Fig. \ref{fig:narrdamas}, by driving AI and/or deception experts from the observation of the phenomenon, to its modelling, analysis and explanation. Representational power would be provided by the ABM and MAS models developed by AI experts. The modularity and scalability of DAMAS would be guaranteed by the MAS engineering of the representations of the world. As for usability, this would be achieved through the user-centric design of the tools focused on the needs of analysts to interact with the MAS models. In the rest of this section we present the components of DAMAS and how these would support the entire deception analysis process. We first describe how deception can be modelled with MAS and ABM approaches drawing from the AI literature in subsection \ref{sec:formsofdec}. Then, in subsection \ref{sec:explain_damas} we discuss how deception can be explained with the aid of user-centric XAI tools based on hybrid argumentation in line with methods from intelligence analysis.

\subsection{Modelling Deception} \label{sec:formsofdec}

The foundations of DAMAS rely on the AI expertise in creating holistic MAS and ABM models of deception from a socio-cognitive perspective. There are a few MAS and ABM approaches in the literature that tackle different forms of deception by considering a multitude of socio-cognitive components. 

Due to their nature, some of these types or forms of deception are more suitable to be modelled and studied at the \textbf{interaction} level (subsection \ref{sec:formsofdec1}) with MAS approaches, while others are more suited for being studied at a \textbf{descriptive} level (subsection \ref{sec:formsofdec2}) with ABM approaches. Henceforth, we will first categorise the forms of deception progressively, starting from the simplest to the more complex and unusual in terms of interaction. Then, we will focus on describing progressively the forms of deception from a descriptive perspective, based on the number of deceptive agents. For both MAS and ABM models we indicate the kinds of deception that have been modelled in the AI literature. We reflect on these models and present alongside some of the questions that remain unanswered in order to guide AI experts into formulating new models or into extending existing ones that would eventually become part of DAMAS.

\subsubsection{Interaction (MAS) Models for DAMAS} \label{sec:formsofdec1}

We start by presenting some forms of deception that take place at the interaction level and that are suitable for modelling with MAS approaches. By adding complex cognitive properties to the MAS models, such as the agents' ability to use Theory of Mind (ToM - see subsection~\ref{sec:psych}), the computational modelling of interactions becomes richer and is able to capture not just what agents communicate between them, but also what goes on inside their minds. Notably, because ToM is an intrinsic component of deception, in this subsection we will look closely at the role ToM plays in each form of deception. More often than not, it is the ToM of the opponent that determines what type of deception takes place at the interaction level, as properties of the ToM that are being used by the agents become the properties of the type of deception employed by the agents, whereas when the interactions are scaled up to a greater number of agents (societies), then the socio-cognitive factors come into play at, which calls for an analysis at the descriptive level. Hence, we suggest the following deception models to be represented and engineered in a multi-layered fashion, and studied at the interaction level within DAMAS.

\paragraph{One-Way Interaction (Deception)} We call a one-way computational deception an interaction between two types of agents: the \textit{deceiver}, Alice, and the \textit{target}, Bob. Alice's goal is to make Bob believe something that Alice believes is false. 

The dynamics Alice and Bob engage in represent the two processes identified by IMT2 as Pars Pro Toto and Totum Ex Parte. As we have mentioned before, deception is about making another infer a false belief, therefore it does not really matter to Alice whether what she tells Bob is true or not as long as Alice manages to make Bob infer a false belief. This here shows the difference between providing false information, which means lying, and providing certain information that leads to a false conclusion, which means deceiving. Depending on the context Alice finds herself in, she needs to decide whether the necessary information she needs to provide Bob in order to deceive him is a lie or a truth, or a half-truth, or a combination of them that is more or less complex. 

A strong focus in AI is on the logical formalisation and categorisation of one-way deception. For example, \citep{sakama2010many} aim to categorise multiple forms of one-way deception and explain their logical properties, and \citep{sakama2010logical} look at the multiple variations of dishonesty including one-way deception, whereas \citep{Caminada-2009-LanBDCofD} looks at the logical relation between truth and other forms of dishonesty such as lying, bullshitting\footnote{Bullshitting, according to \citep{Frankfurt-2009-OnB} is making statements without regard for their truth value.} and even some forms of one-way deception. \citep{sakama2015formal} also uses dynamic epistemic logic to formalise deceptive acts of communication. Later, \citep{sakama2021} adds a causal model for epistemic dynamics determined by communication.

One-way deception has been also studied from an agent-oriented perspective. \citep{derosis2003can} show a simulation of deception in belief networks, whereas \citep{christian2004strategic} define a model of a heuristic plan search for finding a deception plan. More recently,  \citep{PanissonSarkadi-2018-LBandDinAOPL} defined and implemented a BDI agent using Jason (an agent oriented programming language) that can choose to lie, bullshit or deceive in order to manipulate the beliefs of another agent. This work was continued by \citep{sarkadi2019dectom,sarkadi2020phdthesis} who integrated it with TDT, IDT and IMT2. There, the deceptive interactions between two BDI agents are defined and implemented, where the deceiver agent simulates the mind of its target taking into account the levels of the target's trust, the confidence in its ToM of the target, and the level of its communicative skill. The approach introduces the idea that deception can be analysed in multi-agent social interactions not only regarding its success (successful vs unsuccessful), but also regarding whether it was attempted or intended (intended vs unintended).  

A critical question about one-way deception that has not been explicitly explored yet is how would the interaction between a deceiver and its target evolve over time? For example, how would cognitive load and trust increase or decrease and under which conditions? Also, how would the deceiver and the interrogator update their ToM of each other given past interactions? 

\paragraph{Two-Way Interaction (Counter Deception)} - Compared to one-way deception, counter deception eliminates the assumption of Bob's \textit{unknown unknown}. Instead of playing just one role, both Alice and Bob play the roles of the \textit{deceiver} and the \textit{target}. In this case, the same reasoning mechanism is taken to a higher level. Bob's goal is now to deceive Alice into thinking that he has inferred a false belief. Alice's goal is to deceive Bob into thinking he was able to deceive her about deceiving him and so on. The simple fact that Bob is aware of Alice's deceptive intentions, might give away Bob's suspicion. To deceive Alice into thinking he has been deceived, Bob must emulate some behaviour that makes Alice think he was deceived by her. However, if Bob knows that Alice knows that Bob might want to deceive Alice and so on, then what type of behaviour should Bob simulate --- the one indicating that he was deceived or the one indicating that he wasn't deceived --- in order to deceive Alice?

Work in the Intelligence Analysis literature has led to solid psychological theories of counter-deception and deception detection indicating that Intelligence and Espionage Agencies are often engaging in very complex counter-deception analysis \citep{Heuer1999psychology}. Counter deception has also found its applications in interrogations. When interrogators happen to deal with deceptive or manipulative subjects, they can resort to counter-deception to increase their chances at successful interrogation \citep{vrij2012eliciting}. For example, the interrogator can pretend to know some information \textit{a priori} such that the subject is tricked into giving answers that reveal or confirm the truth of the information to the interrogator's questions. Studies show that interrogators trained in counter-deception have a greater success at deception-detection \citep{Hartwig2006strategic}. Moreover, this effect can be also observed in MAS simulations, where deceivers and interrogators seem to engage in a ToM arms race \citep{sarkadi2023arms}.

\paragraph{Two-Way Recursive Interaction} - In theory, counter deception can be  infinitely recursive. The property of recursivity in counter-deception, however, depends on the type of ToM of the opponent. A recursive ToM means that agents add levels of ToM on top of each other's ToM of themselves. For example, ``I know that you know that I know...\textit{ad infinitum}...some information'' represents a recursive ToM. An entirely recursive ToM would mean that the deceptive reasoning processes of a deceiver's mind would only focus on taking a certain belief, let us say $Bel_i(P)$ and infering $Bel_j^k(Bel_i^k(P))$ in order to gain some advantage using deception, where $i$ and $j$ represent different agents and $k$ represents the level of ToM. However, in interactions that assume an entirely recursive ToM, the only advantage belongs to the agent that has a the greater level of ToM as shown by agent-based simulations of games played between agents with multiple orders of ToM \citep{de2017negotiating}.

\paragraph{Partially Recursive Interaction} - In practice, human agents are rationally bounded and are not capable of infinitely recursive reasoning. Thus, in real life, deception is not applied to infinitely recursive mental models. There might be cases in which a deceiver could exploit its target's mind without engaging in expensive recursive reasoning. There might be beliefs that do not exist in the target's ToM of the deceiver. Or there might be beliefs of the deceiver that the target does not know that it (the target) does not know (as in the case of \citeauthor{rumsfeld2002unknown}'s \textit{unknown unknowns}). In such cases, it might be wiser for the deceiver to avoid spending cognitive resources on the higher-order reasoning of ToM and exploit other types of beliefs inside the target's mind. The deceiver might, for example, simulate the belief updates that happen in the mind of the target in order to see what new beliefs can be formed and also explore which of these newly formed beliefs can be more efficiently exploited. This type of ToM implies a dynamic semantic ToM model. Dynamic semantic models of ToM in MAS based on belief-desire-intention architectures and agent oriented communication along with their use under uncertainty have been introduced and implemented by \citep{Sarkadi-2018-UToM}, and more recently described by \citep{shvo2020towards}. One instance of partially recursive counter-deception has also been introduced as preliminary work on argumentation dialogue games, where both the deceiver and the interrogator have a ToM of each other that they update after every interaction, in \citep{sarkadi2019DecStoryAAAI}. The deceiver uses its ToM to build a story that forces the interrogator to accept it, and viceversa, the interrogator forces the deceiver to accept that it has not found a believable story.

\paragraph{Internal Interaction (Self-Deception)} - The exception to the presumably intuitive rule that deception requires at least two agents (deceiver and target) is the case of self-deception. In order for self-deception to be successful, the deceiver must be able to deceive itself, playing both the role of the deceiver and its target. Here we have a paradoxical situation. Assuming that the deceiver needs a ToM of its target in order to deceive, then the deceiver needs a ToM of itself. Given that the same entity plays both the roles, then it must ontologically follow that its ToM of itself must be complete, i.e. there is no knowledge about itself that it does not know. If the ToM is complete, then the ToM must include the deceiver's deceptive intentions or goals as well as the target's goal of not being deceived. Obviously, these two types of conflicting goals and intentions determine an inconsistent system. However, there are some special cases in which these paradoxical situations can be overcome. \citep{Jones2015} manages to model a specific set of cases of self-deception that are logically consistent in Hintikka's logic of belief. The author does emphasise that such inconsistencies still remain inside the system, but become latent due to the specific cases that are formalised, and \citeauthor{Jones2015} explains the reasons for them becoming latent. A fairly new idea has been proposed in by \citep{chendreams2021} to use models of self-deception in designing persuasive software applications to nudge people to sleep better, but even though it is worth mentioning, this type of research is still in its infancy. On a similar note, before \citeauthor{chendreams2021}, \citep{borenstein2016robotic} have actually proposed the idea that ethical robots could be used to nudge humans into making more ethical decisions in their life. Perhaps self-deception can actually drive teamwork and cooperation in hybrid societies? If so, then how do we model interactions to analyse the outcomes of promoting self-deception in humans with AI technologies, or even deploying self-deceptive agents?

\subsubsection{Descriptive (ABM) Models of Hybrid Societies for DAMAS} \label{sec:formsofdec2}

In the previous subsection we have described agent models that deal with different forms of deception at an interaction level. But what happens when we go beyond the mechanics of interaction and try to model deception at a different level of abstraction, e.g. when more than two agents are involved in deceptive interactions that happen as part of a dynamic and large distributed system of agents? To model distributed deception, we need to look at group-based interactions, that are assumed to take place in populations or societies of agents. Here, agents can either rationally decide to change the roles they play based on who they interact with, or they are assigned their roles through some mechanism (usually a stochastic one).

While in the interaction models of deception the focus was on the reasoning and decision mechanisms, here we can assume such mechanisms as a given, and focus on the trade-offs of using combinations of different mechanisms. The trade-off itself should depend on the factors that influence deception such as the types of other agents they interact with, the information available to them, their available ToMs, the cognitive load of the agents, their communicative skill, the trust between them, the communication protocol they follow (or the specific game they play). Depending on the type of each distributed system the agents belong to, the cost of deceiving, interrogating or counter-deceiving might differ. Therefore, an overarching research question for distributed deception would be how does the cost of deception influence agents in large scale interactions? We classify distributed deception in three types to see what other relevant questions might drive its agent-based modelling with DAMAS.

\paragraph{Type I: Multiple deceivers and a single target} - The obvious problem would be for deceivers to find a way in which they are able to maximise the likelihood of their success through cooperating with each other. How do they cooperate with each other to deceive their target, assuming that all of the deceivers share a single goal in terms of what false belief they want their target to infer? More specifically, how do the deceivers manage to execute \textit{Pars Pro Toto} efficiently between themselves?  What information do they have to distribute between themselves, what information does each of the deceivers have to withhold and which information does each of them have to send and in what way? Which of them has to lie and which has to tell the truth, and in what sequence? Assuming they require a ToM to deceive, do they have to share information about their ToMs of the target between themselves? Does the presence of more deceivers mean a higher likelihood of success, or does it hinder the deceptive process by adding layers of reasoning and increase the complexity of reasoning and decision making?

\paragraph{Type II: A single deceiver and multiple targets} - A single deceiver has to account for multiple targets. This case should not be confused with multiple one-way deception where a single deceiver repeatedly employs one-way deception separately for each target. In type II, the deceiver needs to take into account at least more than one target when attempting a deceptive act. One research question would be how does the deceiver disseminate information to its different targets given different combinations of constraints? The targets of the deceiver might cooperate by sharing and comparing information between themselves in order to protect themselves from deceivers. Maybe only some of them cooperate and some of them do not. Perhaps the deceiver has multiple targets, but only one of them is crucial for its success. Thus, another question might be how can the deceiver exploit cooperation and non-cooperation between its targets in order to successfully deceive? 

\paragraph{Type III: Multiple deceivers and multiple targets} - It might be interesting to assume that agents are able to play both the role of deceivers and potential targets. Agents are able to decide what role to play by calculating their pay-offs to see whether is it profitable (rational) to a) deceive a target or b) to risk being a target itself and blindly trust the agent it interacts with or c) try to only act as a target in order to interrogate or counter-deceive the other agent. Moreover, is there a different pay-off when trying to deceive more than one agent? What if the deceiver needs to interact with multiple agents at the same time or in a certain given sequence? Are all of these agents easy targets, or are some of them counter-deceivers? Finally, how do we control or govern a society where this type of deception can happen?

Regarding the related work in MAS and ABM on distributed deception as we have defined it here, the closest works would be the one on evolution of deception in hybrid societies by \citep{sarkadi2021evolution,sarkadi2024triangles} who model large-scale and long-term interactions as public-goods-games, and the work on the profitability of incompetence by \citep{staab2010profitability} who define artificial agents that bullshit their way through society in order to maintain the view that they (the agents) are competent.

\bigskip

To summarise, MAS approaches can be used to model deception at the interaction level, e.g. what happens inside the agents' minds when they communicate with each other - internal cognitive processes and architectures responsible for deceptive communication. On the other hand, ABM models can be used to model distributed deception, e.g. to describe the states and equilibria given the emergent behaviour of agents influenced by the socio-cognitive factors of societies where deception is possible. However, simply having rich agent-based models is not enough for non-experts (policy-makers, regulators etc.) to understand deception, and more importantly, to make informed decisions regarding deceptive technologies in different contexts. Human analysts need to make sense of these models and need to know how to use them in order to methodically study deception in a critical manner. This is where the literature and methods from Intelligence Analysis could lend the AI community a helping hand in designing a MAS framework that is scientifically non-reductionistic, usable, and explainable.


\begin{table}[]
    \centering
    \begin{tabular}{|l|p{10cm}|}
    \toprule
    \textbf{Model} & \textbf{Some DAMAS-compatible approaches} \\
    \midrule

        One-Way Deception & \citep{sakama2010many,sakama2010logical,Caminada-2009-LanBDCofD,sakama2015formalIGPL,derosis2003can,christian2004strategic,PanissonSarkadi-2018-LBandDinAOPL,sarkadi2019dectom}\\
        
        \hline
        
        Counter-Deception & \citep{sarkadi2019DecStoryAAAI}* \\
        
        \hline
        
        Self-Deception & \citep{Jones2015}, \citep{chendreams2021}* \\
        
        \hline
        
        Distributed Deception & \citep{staab2010profitability,sarkadi2021evolution,sarkadi2024triangles}\\
        
        \bottomrule
    \end{tabular}
    \caption{The four types of agent-oriented deception and examples of papers that describe AI approaches which can be used to model the respective types. The notation * represents preliminary work.}
    \label{tab:dectaxonomy}
\end{table}
 
\subsection{Explaining Deception - Hybrid Argumentation Tools for DAMAS} \label{sec:explain_damas}

For the interaction and descriptive models of DAMAS to be used in real-world analysis and decision making - in a meaningful manner - DAMAS should provide analysts (end users, not just AI scientists and experts) with some means or methods to explain and interpret the ABM and MAS dynamics. Remember that in history and Intelligence Analysis, to understand deception in a meaningful manner, deception causation and causation prevention need to be understood as part of an overarching narrative~\citep{ferris1989intelligence}. 

In deception analysis, counterfactual or hypothetical reasoning deals with questions that are relevant for establishing event causation and event causation prevention. A promising approach in AI to address these questions is to use argumentation-enabled tools, which could be integrated in DAMAS for automated reasoning. Cybersecurity research is already looking at how the area of argumentation can be used to perform this type of analysis for tasks such as cyber attribution \citep{simari2019data}.  Argumentation-based approaches can also be used in criminal forensics as proposed by \citep{bex2010story} and backed by \citep{zenker2019stories}. The implementation of such support tools for intelligence analysis is not new and is indeed feasible, as previous work by \citep{toniolo2015supporting} demonstrates. 

Given these capabilities of argumentation-enabled tools to facilitate explanations, we recommend to include in DAMAS \citeauthor{bex2011arguments}'s hybrid approach to argumentation and storytelling to explain causation and causation prevention as a narrative such that it is intelligible for analysts (see Fig. \ref{fig:narrdamas}). \citep{bex2011arguments}'s approach has been designed to both represent causal chains of a main story and to use arguments to anchor the main story's sub-stories in evidence, a process named \textit{anchoring} which results in explanations. By applying this approach, one does not only find an arbitrary story for a causal chain of events, but is able to select the best main story (out of several viable ones) that is composed of several sub-stories, e.g., perform the inference to the best explanation (IBE), by explaining through arguments how the main story's sub-stories are backed by evidence.

DAMAS's rich socio-cognitive representations of deception dynamics, provided by the framework's representational power, would allow the extraction of complex chains of stories where multiple agents, events, and interactions between agents take place. By being able to extract and interpret these complex chains of events, intelligence analysts would be able to reason more critically and more clearly about deceptive scenarios, by considering multiple configurations of agents at different levels of abstraction. In turn, this would allow analysts to contrast and compare multiple narratives and select the one with the strongest explanatory power - in other words, to perform IBE. 

Representing multiple levels of abstraction should help analysts take into account the complete and partial knowledge of the agents involved in the respective scenarios - interaction MAS level, as well as the effect of the social factors that influence deception parameters - descriptive ABM level. Subsequently, analysts could describe and contrast the cognitive and communicative differences, while comparing similarities, that follow from the representations and behaviour of deceptive and non-deceptive agents. For example, it should be possible for an analyst to distinguish a lie from a lie with deceptive intent \citep{chisholm1977intent}, if components of deception could be described by the DAMAS modules by taking into account relevant theories and taxonomies of deceptive communication. Additionally, it is crucial for the analysts to be able to say if there is any intentional, unintentional, or even self-deception happening \citep{whaley1982toward}, and be able to identify whether or not `banal' deception is the case - deception that arises from using a technology that does not deceive on its own, but it is designed to be deceptive in a certain context \citep{natale2021deceitful}. A good way to identify banal deception in hybrid societies would be to look at cases where deception is caused by semi-autonomous machines which cannot be attributed intention \citep{petasurvey2021}, and represent these in DAMAS.

Using Bex's approach, the observable behaviour of agents inside DAMAS could be used to test whether a potential viable \textit{main story} (a complex hypothesis) about their behaviour counts as an explanation of their behaviour. How well the story is anchored in the evidence could also be observed in DAMAS, e.g., how many items of evidence confirm or falsify the story. Furthermore, the story could be used to confirm or falsify how certain events in the real world might have unfolded, but more importantly it could explain how they might have unfolded or not if something else were the case. These are precisely the sorts of explanations of agent behaviour that human minds try to figure out in socio-cognitive interactions \citep{malle2006mind}. One must also be careful about grounding the representation in relevant theories, as selecting bias-prone theories for this purpose leads to errors in analysing deception using AI methods \citep{sanchez2020politics,stark2021physiognomic}.
Anyways, \citeauthor{bex2011arguments}'s hybrid approach could also be used to find biases and inconsistencies inside DAMAS w.r.t. the theories that have been used to implement the MAS models. Perhaps there is considerable evidence outside the MAS that indicates that the agents' behaviour should have been different. DAMAS would allow then the identification of the underlying cause for the unexpected behaviour, at which level of abstraction it happens, and whether it could be calibrated inside the model that represents that particular level of abstraction. Subsequently, following the feedback of analysts, AI experts could directly revise the underlying AI model according to a new evidence-based theory and re-implement it inside the framework. This process would be facilitated by DAMAS's scalability and modularity.  

Stories can change based on new evidence, new characters, or new rules, which must be considered by DAMAS. A story can represent how a new agent character enters the story, or how the chain of events moves from one location (domain) to another, or perhaps the narrative style of the story allows for referring back to evidence or arguments that have a certain temporal property (using information available at a different time), e.g. something happened in the past that affected the current chain of events, or perhaps if this chain of events continues in this direction, something relevant to the story will happen in the future. Again, this would be facilitated by following the principles of scalability and modularity of designing DAMAS models in tune with the open nature of MAS systems~\citep{van1999open,pinyol2013computational} where new types of agents or software tools are added or removed from the system.

By representing dynamics at different layers of abstraction in the world models, DAMAS could enable the generation of narratives from multiple points of view (PoV). In this way, one could explain a phenomenon from the PoV of a single agent, from the PoV of multiple agents that can either have been actively involved in a context or have only been passive observers, or perhaps from the PoV of a Big Brother observer, one that has access to multiple viewpoints of agents, as well as extensive knowledge of the chains of events and the domains in which these events have happened. Hence, to do this, DAMAS should have models that account for (i) \textit{spatial scalability}, such as the analysis of deception on large scales, e.g., in various regulatory systems, organisations and societies of different sizes where agents can interact; and for (ii) \textit{temporal scalability}, e.g., changes and evolution of interactions between agents over time. 

To convey these complex narratives to users and to assist with the development of XAI tools and models, socially interactive AI agents can be designed as part of DAMAS to promote usability. When building DAMAS XAI tools based on hybrid argumentation, it is desirable to go beyond the identification of causal relations and try to embed mechanisms that generate user-friendly explanations of deceptive interactions. But what is a user-friendly explanation?

According to \citep{miller2018explanation}, explanations do not consist solely of using abductive reasoning to find causal relations, but they must also include a social process, which refers to the way of convening the knowledge to be exchanged between explainer and explainee. Regarding MAS tools, socially-aware AI meta-agents can be modelled and engineered as part of DAMAS so that they can automatically narrate various chains of complex events to analysts. DAMAS models that enhance abductive reasoning should be designed as to enable these socially-aware AI meta-agents to extract narratives directly from the agent dynamics of the MAS. Adding argumentation capabilities to socially-aware meta-agents could help analysts also engage in critical and hypothetical thinking about possible narratives using question-answering games. It is here where research in dialogue argumentation games for agent social interaction can provide mechanisms for reaching sound conclusions \citep{McBurney-2009-DGforAA}. The cooperation between humans and artificial agents would result in a better understanding of deceptive scenarios through dialectical reasoning. For instance, the approaches described in \citep{shvo2020towards,Panisson-2018-SforToM,Sarkadi-2018-UToM,mosca2021elvira,mosca2020agent} could be combined to model the social process between artificial storytelling meta-agents and their interlocutors (explainees). These approaches seem appropriate, because most humans prefer to explain agent behaviour in a practical reasoning manner by referring to beliefs, desires, intentions etc. \citep{malle1999people,malle2006mind}. Storytelling meta-agents could model the minds of the explainees in order to narrate complex chains of events in an efficient manner - achieving the goal of explaining the phenomenon itself by taking into account the social context in which they explain the phenomenon. Conclusively, by interacting with a usable DAMAS, the analysts' workload and tedious tasks, such as manually performing SATs, would be reduced and simplified through the interactions with socially-explainable AI agents.

To summarise, in order to support deception analysis, DAMAS should be built starting from existing AI techniques that consider the need of intelligence analysts to understand event causation and event causation prevention as part of overarching narratives. One approach to perform deception analysis with DAMAS in this manner is to combine AI models and tools with Bex's hybrid approach of using stories to perform IBE of complex cases. Together with rich representations, modularity, and scalability of agent dynamics in DAMAS, multiple narratives could be extracted from agent dynamics to directly perform IBE for different complex and open scenarios. Finally, if analysts, given their expert knowledge about the real world, believe there is an alternative best explanation that can be inferred outside the MAS models, they should be able to interact with AI experts in order to refine the models within DAMAS. In this way, analysts would be able to discover new agent dynamics and further test their hypotheses from a socio-cognitive perspective, supported by an ever-improving AI framework for deception modelling and analysis.

\section{Conclusion}

The future of hybrid societies may be threatened by the development of deceptive AI. To understand the effects of deception between agents of hybrid societies, whether these agents are humans or machines, in this paper we proposed the development of DAMAS. DAMAS is envisioned as a multi-agent framework for enhancing the socio-cognitive modelling, explanation, and analysis of deception from the perspective of Intelligence Analysis

We started the paper by introducing two important perspectives on deception, namely the philosophy of deception and a socio-cognitive perspective. The philosophical perspective aimed to define and explain what deception is and what it is not, by clarifying subtle distinctions in the literature. The role of the socio-cognitive perspective was to familiarise the readers with the socio-cognitive factors and components responsible for deception in humans.

Next, we argued that if we aim to model deception with AI techniques, then we must adopt a holistic socio-cognitive perspective because otherwise we would fall into the trap of scientific reductionism. We further argued that reductionism in AI is not in itself bad, but that in order to build a holistic understanding, then the existing AI approaches must be integrated as part of an overarching MAS framework. Our argument echoes what has been proposed about 20 years ago regarding the development of a theory to understand the role of trust and deception in virtual societies, notably by  \citep{castelfranchi2000artificial,castelfranchi2001trust,castelfranchi2002role}. The reason we have reinforced this perspective is because despite the significant number of contributions on socio-cognitive modelling of trust, the socio-cognitive modelling of deception has not had the same momentum. Having surveyed the literature on deception modelling in AI, we believe it is now the time for the next stage in deception modelling and analysis.

Hence, to take the modelling of deception to the next stage, we proposed DAMAS. We described its potential to bridge the scientific modelling and engineering of deception with AI techniques with the methods and techniques used by intelligence analysts to understand and explain deception in real-world complex cases. If developed, DAMAS would be able to act as a platform that enables (i) AI experts to formulate approaches for the socio-cognitive modelling of deception; (ii) AI experts and/or engineers to represent and implement these models in multi-agent frameworks at multiple layers of abstraction (environment, agent, organisation);  and (iii) intelligence analysts to use XAI hybrid argumentation tools to extract and analyse agent-based narratives in order to perform inference to the best explanation (IBE).

Finally, we discussed the impact of being able to understand deception holistically in hybrid societies and what are the ethical considerations of modelling deceptive AI. We pointed out the risk of reaching a Tragedy of the Digital Commons \citep{greco2004tragedy} and discussed the need of developing accountable, responsible and transparent deceptive AI w.r.t. ethical guidelines, such as the guidelines of HLEG for trustworthy AI \citep{hleg2019ethics} and the ones of IEEE on Ethically Aligned Design \citep{chatila2019ieee}. We then presented two arguments that defend the modelling and development of trustworthy deceptive AI for the benefit of society under the condition of following ethical principles of such as value-alignment, transparency, and scientific discovery.

To conclude, we wish to emphasise that deceptive AI has both drawbacks and benefits and that to be able to understand what these are and reason about them, a holistic socio-cognitive approach to modelling deception in hybrid societies is necessary. We hope that the development of DAMAS will be the next step in achieving this understanding such that future hybrid societies will be able to reap the benefits of trustworthy deceptive AI and avoid the risks associated with the future development and deployment of malicious deceptive AI.

\bibliographystyle{unsrtnat}  
\bibliography{references}

\end{document}